\documentclass[twocolumn]{aastex63}

\usepackage{commath}

\newcommand{\Ms}{{\ensuremath{M_{\odot} }}}

\accepted{ApJL}

\shorttitle{Binary DCBHs }
\shortauthors{Latif et al.}
\begin{document}

\title{The Birth of Binary Direct-Collapse Black Holes}

%% you desire. In v6.3 you can include a footnote in the title.

%% A significant change from earlier AASTEX versions is in the structure for 
%% calling author and affiliations. The change was necessary to implement 
%% auto-indexing of affiliations which prior was a manual process that could 
%% easily be tedious in large author manuscripts.
%%
%% The \author command is the same as before except it now takes an optional
%% argument which is the 16 digit ORCID. The syntax is:
%% \author[xxxx-xxxx-xxxx-xxxx]{Author Name}
%%
%% This will hyperlink the author name to the author's ORCID page. Note that
%% during compilation, LaTeX will do some limited checking of the format of
%% the ID to make sure it is valid. If the "orcid-ID.png" image file is 
%% present or in the LaTeX pathway, the OrcID icon will appear next to
%% the authors name.
%%
%% Use \affiliation for affiliation information. The old \affil is now aliased
%% to \affiliation. AASTeX v6.3 will automatically index these in the header.
%% When a duplicate is found its index will be the same as its previous entry.
%%

\correspondingauthor{Muhammad A. Latif}
\email{latifne@gmail.com}

\author{Muhammad A. Latif}
\affiliation{Physics Department, College of Science, United Arab Emirates University, PO Box 15551, Al-Ain, UAE}

\author{Sadegh Khochfar}
\affiliation{Institute for Astronomy, University of Edinburgh, Royal Observatory, Blackford Hill, Edinburgh EH9 3HJ, UK}

\author{Daniel Whalen}
\affiliation{Institute of Cosmology and Gravitation, University of Portsmouth, Portsmouth PO1 3FX, UK}
\affiliation{Ida Pfeiffer Professor, University of Vienna, Department of Astrophysics, Tuerkenschanzstrasse 17, 1180, Vienna, Austria}

\begin{abstract}

Supermassive primordial stars forming during catastrophic baryon collapse in atomically-cooling halos at $z \sim$ 15 - 20 may be the origin of the first quasars in the universe.  However, no simulation to date has followed the evolution of these halos at resolutions that are high enough or for times that are long enough to determine if collapse actually produces SMSs.  Here we report new cosmological simulations of baryon collapse in atomically-cooled halos for times that are long enough for SMSs to form and die as direct-collapse black holes (DCBHs).  We find that the high infall rates required to build up such stars do persist until the end of their lives and could fuel the rapid growth of their BHs thereafter.  Our simulations also demonstrate  that binary and even small multiples of SMSs can form in low-spin and high-spin halos, respectively.  This discovery raises the exciting possibility of detecting gravitational waves from DCBH mergers with LISA and tidal disruption events in the near infrared with the {\em James Webb Space Telescope} and ground-based telescopes in the coming decade.
% for the first time
\end{abstract}


\section{Introduction} \label{sec:intro}

More than 300 quasars have now been discovered at $z > 6$ \citep[e.g.,][]{Fan2003}, including seven at $z > 7$ \citep{Mortlock2011,Banados18,Matsuoka19}.  The formation of such massive black holes less than a Gyr after the Big Bang poses serious challenges for paradigms of early structure formation.  A number of processes have been proposed for the origins of these quasars: the collapse of Pop III stars to BHs at $z \sim$ 20 - 25, runaway collisions in dense nuclear clusters at $z \sim$ 10 - 20 that build up a single massive star that collapses to a black hole, and the formation of supermassive stars in atomically-cooling halos at $z \sim$ 15 - 20 that die as direct-collapse black holes (DCBHs). 

Pop III remannant BHs are expected to range in mass from a few tens to hundreds of solar masses at birth \citep[e.g.,][]{Hirano2014} but are born in low ambient densities that preclude their initial rapid growth \citep{Whalen04,Johnson2007}, and some are ejected from their halos at high velocities by natal kicks \citep{wf12}.  They in principle could reach 10$^9$ \Ms\ by $z \sim$ 7 with episodes of super-Eddington accretion at duty cycles of just a few percent \citep{Pezzulli16,Lupi16} but have not encountered high enough densities to trigger such growth in any cosmological simulation to date \citep{Alvarez2009,srd18}.  Runaway stellar collisions in marginally-enriched dense stellar clusters can create BHs of up to a few thousand solar masses \citep{Devecchi2012,Latif2016dust,Reinoso18a} but even these objects may not be massive enough to become quasars by $z >$ 6 \citep{Smidt18}.

For these reasons DCBHs have become the leading contenders for the seeds of the first quasars.  Catastrophic baryon collapse in atomically-cooling halos leads to initial infall rates of $\sim$ 0.01 - 1 \Ms\ yr$^{-1}$.  Standalone models of stellar evolution have shown that if such rates persist they would build up cool, red 100,000 - 300,000 \Ms\  stars before they collapse to DCBHs via  general relativistic instability \citep{Hosokawa2013,um16,hle17,hle18}.  The low ionizing UV fluxes of these stars cannot slow down accretion onto themselves so DCBHs form in the dense environments that create them and can therefore grow much faster at birth.  But for these halos to reach masses of 10$^7$ - 10$^8$ \Ms\ and virial temperatures of $\sim$ 10$^4$ K without having first formed a less massive Pop III star via H$_2$ cooling, they must grow either in the presence of Lyman-Werner (LW) UV sources that sterilize them of H$_2$ \citep[e.g.,][]{Sugimura14,Latif2015a,agarw15} or in highly supersonic baryon streaming motions that delay the collapse of the halo even if H$_2$ is present \citep{srg17,Hirano17}.  Once collapse begins it proceeds quickly because of high densities, temperatures, and therefore sound speeds, $c_{\rm s}$, in the gas ($\rm \dot{M} \sim {c_s^3}/{G} \sim 0.1~M_{\odot}/yr \left( {T}/{8000~K}\right)^{3/2}$).

Numerical simulations of the collapse of atomically-cooling halos at high redshifts have steadily improved over the past decade but remain a trade-off between resolution and evolution time. The original simulations either resolved sub-AU scales that could only follow the formation of the hydrostatic protostar \citep[but not the atomically-cooled disk around it;][]{Wise2008} or 0.01 pc scales that captured the formation of the disk but still could only follow its evolution for a few dynamical times \citep{Regan09}.  These studies found large infall rates at early times but could not determine how long they lasted.  Later work at high resolution and somewhat longer evolution times found that large accretion rates continued down to scales approaching those of the supermassive star itself but did not run for nearly enough times to follow its evolution \citep{Latif2013a,Regan2014a,Latif2016}.  

The introduction of sink particles and pressure floors \citep{Machacek01} at the highest resolutions extended simulation times to a few tens of thousands of years \citep{Latif2013d,Shlosman2016,Regan18b,Becerra18}, confirming that accretion rates at the smallest scales remained high.  Most recently, \citet{Chon18} and  \cite{Regan18b} employed sink particles with radiative feedback from the protostar with a simple treatment of its evolution to follow collapse for 100 kyr and 250 kyr, respectively.  They confirmed that radiation from the protostar was unable to prevent high accretion rates and found some small scale fragmentation in the disk at later times.  In particular, \citet{Chon18} found that almost half of the fragments in one of the simulated halos are formed in binaries that they suspected may become supermassive binaries, corroborating idealized simulations by \citet{Bromm03}.  However neither study evolved the disks for long enough times to determine if any of the fragments became stars or were simply subsumed into the central object at later times.  Previous work also only considered one or a few halos that were not parametrized by either spin parameter or assembly history so how these factors influence the masses or numbers of supermassive stars in the halos remains unknown.

We here follow the collapse of atomically cooled halos at high redshift for time scales 12 times longer than in any comparable simulation to determine the final fates of any fragments that form in their accretion disks, how many supermassive stars might result, and their masses at collapse to DCBHs.  Our halos were chosen to have a range of spin parameters that bracket their likely values at early epochs and we have tallied accretion rates at the centers of the disks for later use in supermassive stellar evolution models in the cosmological flows that create the stars.  In Section 2 we describe our numerical methods and simulations and discuss our results in Section 3.  We examine the consequences of our results and conclude in Section 4.

\section{Numerical Method} \label{sec:methods}

We model the collapse of atomically-cooled halos with the Enzo adaptive mesh refinement cosmology code \citep{Enzo2014}.  Enzo utilizes an $N-$body adaptive particle-mesh scheme to evolve dark matter (DM), a 3rd-order piecewise-parabolic method for fluid dynamics, a multigrid Poisson solver for calculating self-gravity, and a nonequilibrium reaction network to evolve primordial gas chemistry \citep{Anninos1997}.  Simulations are initialized with Gaussian primordial density fluctuations at $z =$ 150 with MUSIC \citep{Hahn2011} with cosmological parameters from the second-year \textit{Planck} best fit lowP+lensing+BAO+JLA+H$_0$: $\Omega_{\mathrm{M}}=$ 0.308, $\Omega_{\Lambda}=$ 0.691, $\Omega_{\mathrm{b}} = $ 0.0223, $h =$ 0.677, $\sigma_8 = $ 0.816, and $n =$ 0.968 \citep{Planck2016}.  To approximate the presence of a strong LW background we only evolve $\rm H, ~H^+, ~He,~ He^+, ~He^{++} and ~e^-$ and neglect H$_2$ and HD chemistry.

\begin{figure*} 
\begin{center}
\includegraphics[scale=0.8]{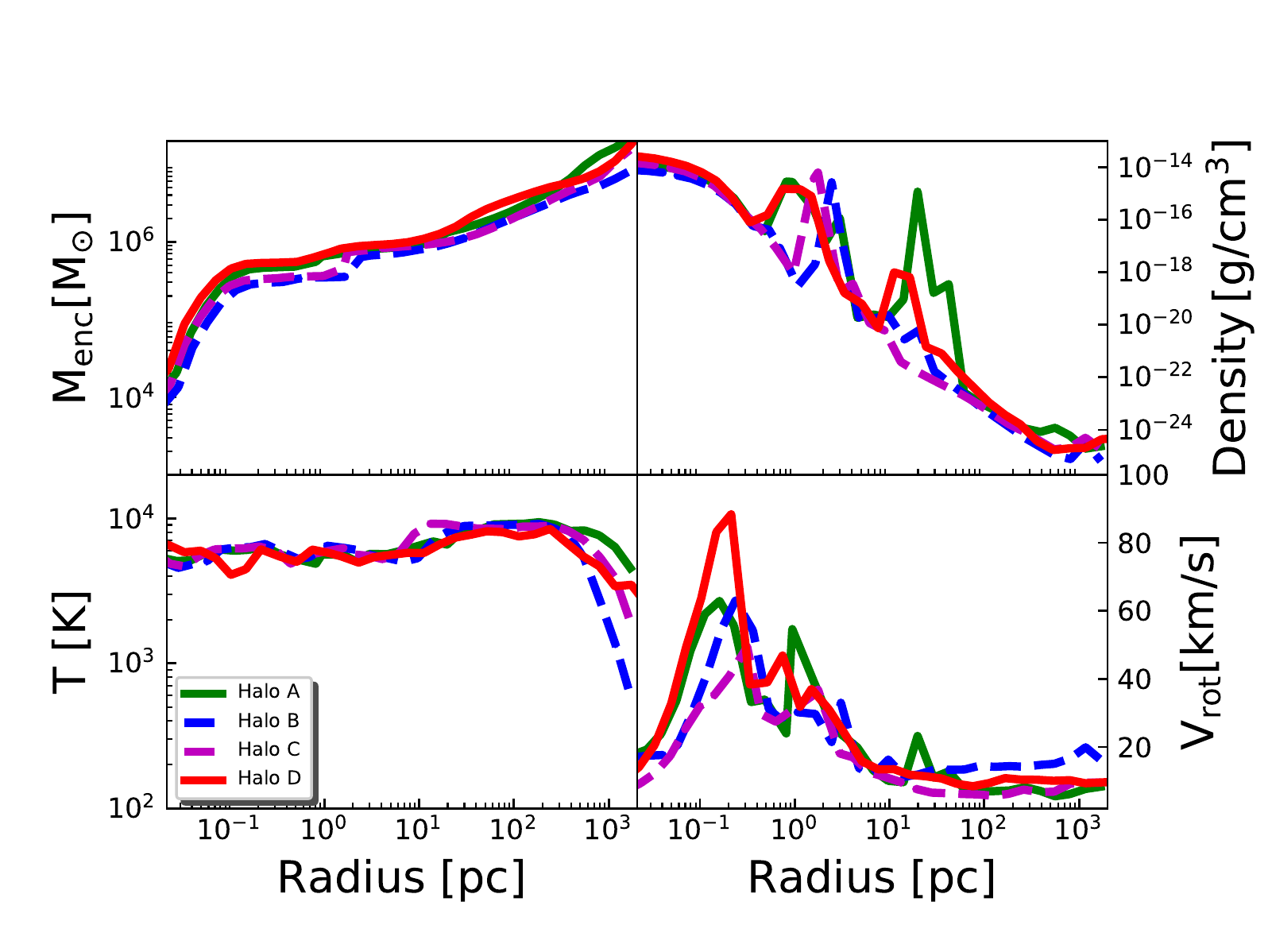}
\end{center}
\caption{Spherically averaged profiles of enclosed gas mass, density, temperature and rotational velocity 3 Myr after the onset of collapse. Green: halo A; blue: halo B; magenta: halo C; red: halo D.}
\label{fig1}
\end{figure*}

\begin{figure*} 
\begin{center}
\includegraphics[scale=0.4]{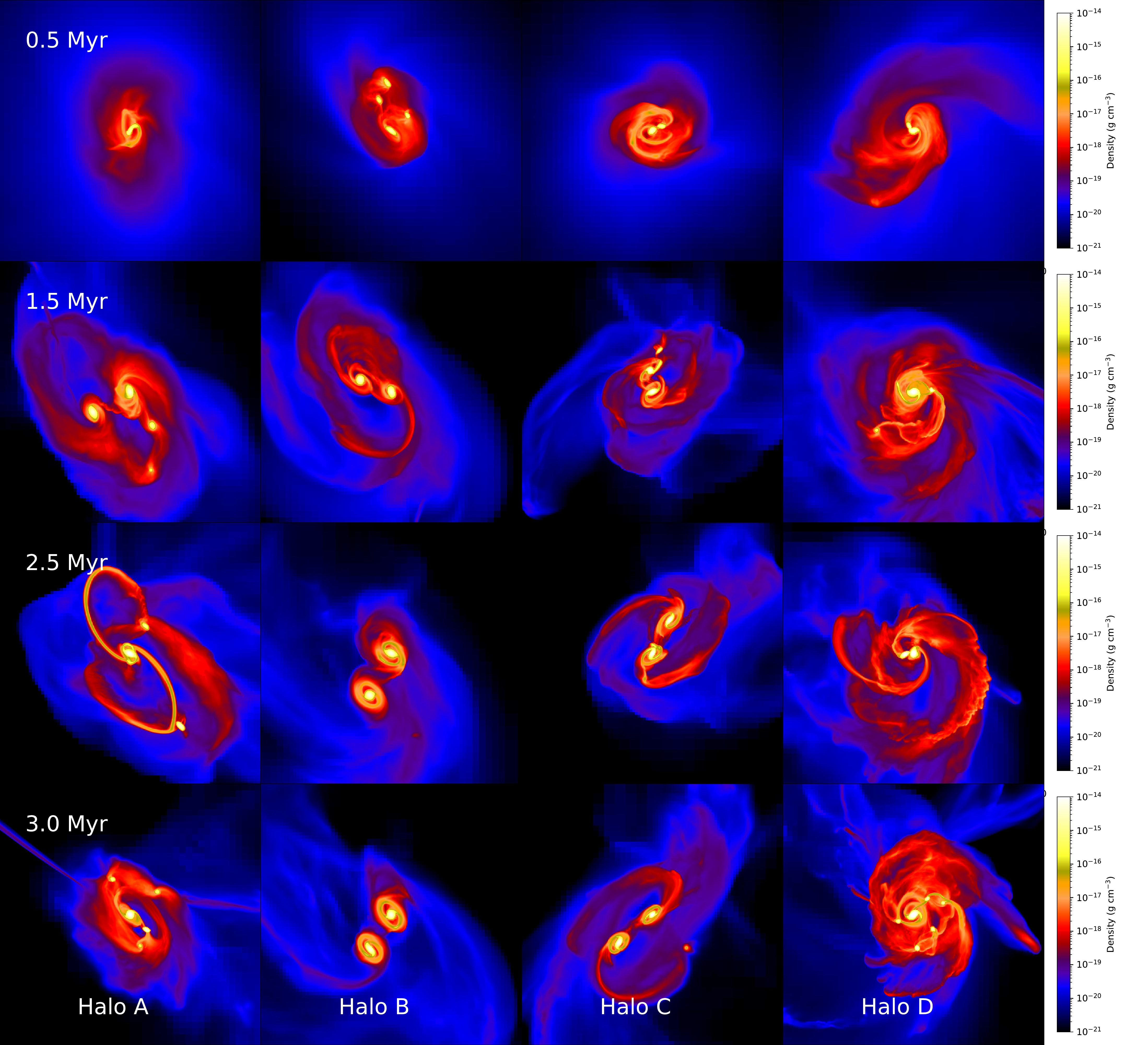}
\end{center}
\caption{Projections of gas density for the disks in our runs.  Columns 1 - 4 are halos A - D, respectively, and rows 1 - 4 are 0.5 Myr, 1.5 Myr, 2.5 Myr and 3 Myr.  Each image is 10 pc on a side.}
\label{fig2}
\end{figure*}

We use  L = 1 cMpc h$^{-1}$ simulation box with periodic boundary conditions and run a number of DM-only simulations at low resolution to identify random seeds that produce halos that exceed 10$^7$ \Ms\ at $z > 10$ with a variety of spin parameters $\lambda = J \abs{E}^{1/2}/G M^{5/2}$, where $J$ is the total angular momentum, $E$ is the total energy (kinetic plus gravitational), and $M$ is the total halo mass.  The halos we chose have $\lambda=$ 0.08, 0.005, 0.002 and 0.02 (labeled A, B, C \& D, respectively), which span the spin parameters found in cosmological simulations \citep[e.g.,][]{Bullock01}). We considered this range of $\lambda$ to investigate its impact on DCBH formation.  In each simulation we center three static nested grids on the halo, a top grid and two additional grids enclosing the central 20\% of the top grid with the same resolution ($\rm 256^3$) and number of DM particles.  This setup yields an initial effective resolution of $\rm 1024^3$.  

We allow up to 15 levels of refinement during the simulation for a maximum spatial resolution of $\sim$ 2000 AU and minimum DM particle mass of $\sim$ 67 \Ms. The grid is refined on baryon overdensity and DM particle mass, and we ensure that the Jeans length is resolved by at least 64 cells during the run, which has been found to be sufficient to resolve turbulent eddies in past cosmological simulations \citep{Latif2013}.  We employ a pressure floor to stabilize collapse on the smallest scales after reaching the maximum refinement level. This approach enables us to follow the evolution of clumps that could become SMSs out to 3 Myr.  Further details on our simulation setup and refinement criteria can be found in \cite{Latif2016} and \cite{Latif19}. 

\section{Results} \label{sec:results}

\begin{figure} 
\plotone{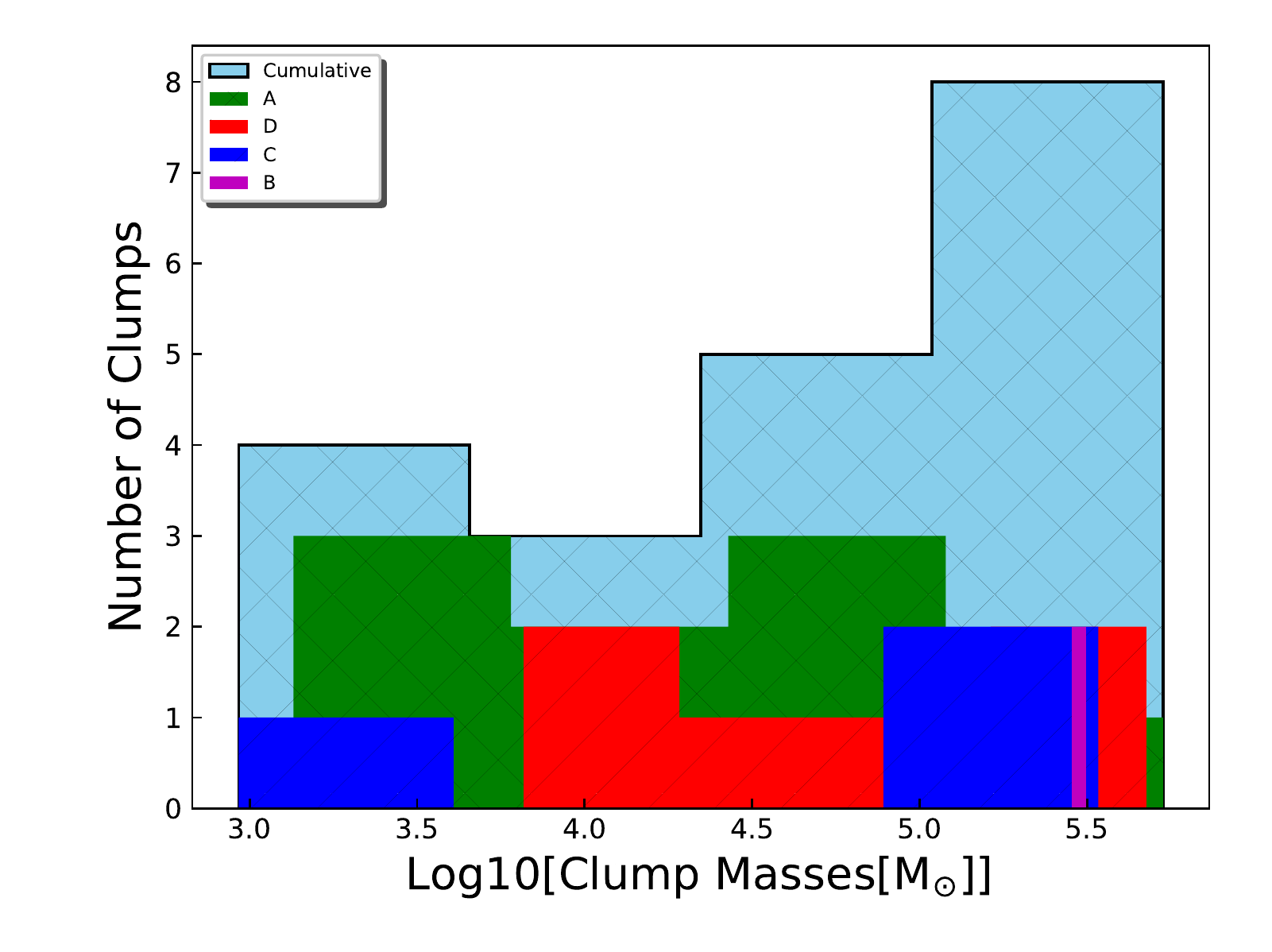}
\caption{Mass distribution of clumps for all four halos along with cumulative clump mass at 3 Myr.}
\label{fig3}
\end{figure}

Halos A, B, C and D begin to atomically cool at masses of $5.16 \times 10^{7}$ \Ms, $1.8 \times 10^{7}$ \Ms, $2.1 \times 10^{7}$ \Ms\ and $2.6 \times 10^{7}$ \Ms\ at $z = $ 10.8, 10.3, 12.7, and 13.3, respectively.  We evolve each halo for 3 Myr after the onset of collapse (point when maximum refinement level is reached in simulations with densities of $\sim$ 10$^{-18}$ g cm$^{-3}$) because this is enough time for an SMS to form and collapse to a DCBH at the infall rates we find in the accretion disks in our simulations, as we discuss below \citep[see also Figure~4 of][]{tyr17}.  Spherically-averaged profiles of all four halos at 3 Myr are shown in Figure~\ref{fig1} and projections of the accretion disks forming in them are shown at 0.5 Myr, 1.5 Myr, 2.5 Myr and 3 Myr in Figure~\ref{fig2}.  They collapse nearly isothermally with central temperatures of $\sim$ 5000 - 8000 K.  The halos have similar profiles at larger radii because of the self-similar nature of runaway gravitational collapse.  Gas in the halos at radii greater than 10 pc exhibit Keplerian rotation with velocities of $\sim$ 15 km s$^{-1}$.  At smaller radii (between ~0.01 - 0.1 pc) these velocities rise sharply because of the formation of self-gravitating disks, as seen in Figure~\ref{fig2}.

Gas densities vary from $\sim$ 10$^{-24}$ g cm$^{-3}$ at the virial radii (several hundred pc) to $\sim$ 10$^{-14}$ g cm$^{-3}$ at the center, and have roughly the $r^{-2}$ profiles expected for isothermal collapse.  The bumps in the density profiles are due either to massive clumps within disks, as in halos A and D, or to binary disks at a few pc, as in halos B and C.  The enclosed gas mass increases sharply outward from the center and flattens out further out in the disks, which have masses of a few 10$^5$ \Ms.  The disks in the high spin halos A and D are more massive than the those in halos B and C.  The enclosed gas mass within the virial radius reaches few times $\rm 10^6~M_{\odot}$ in all four cases.

As shown in Figure~\ref{fig2}, solitary disks form in halos A and D and partially fragment into a few satellite clumps, but halos B and C form binary accretion disks.  Fragmentation is more frequent in the solitary disks, with clumps tidally stripping mass from each other at times (as shown, for example, at 2.5 Myr in halo A).  At 1 Myr these fragments have typical masses of a few 10$^4$ \Ms.  Most of them spiral into the massive clump at the center of the disk well before they could form stars but a few are ejected into the surrounding medium and survive for 2 Myr.  The accretion disks in halos B and C have masses of 2 - 3 $\times$ 10$^5$ \Ms\ at 1 Myr.  Although they also exert tidal forces on each other, they survive for 2 Myr, with average separations of 2 pc.  The numbers and masses of the clumps in all four halos at 3 Myr are shown in Figure~\ref{fig3}.  The most massive ones reach a few 10$^5$ \Ms, with a few $10^3 - 10^4~M_{\odot}$ fragments in the high-spin halos.    

\begin{figure*}
\hspace{-6.0cm}
\centering
\begin{tabular}{c c}
\begin{minipage}{6cm}
\vspace{-0.2cm}
\includegraphics[scale=0.5]{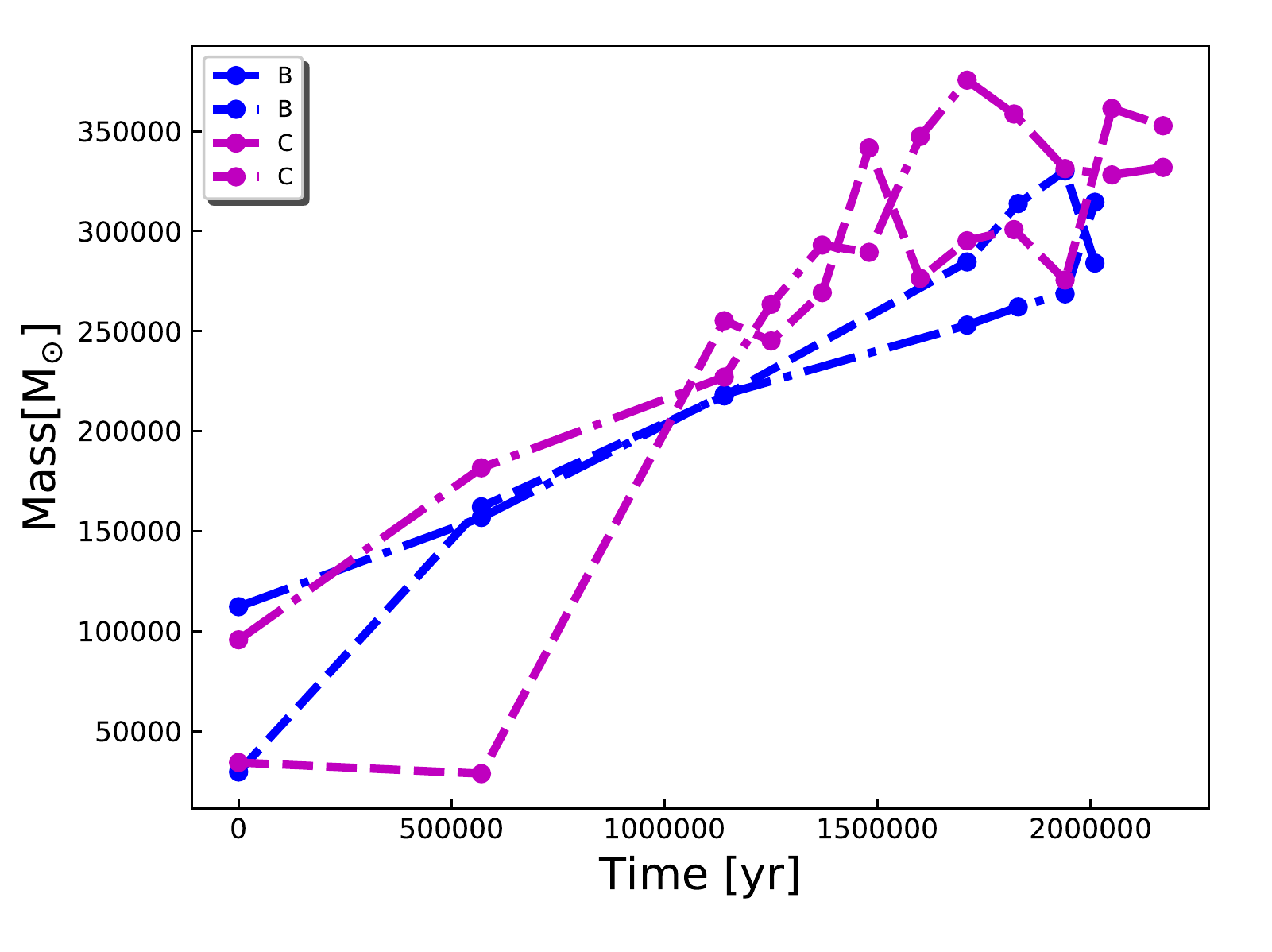}
\end{minipage} &
\begin{minipage}{6cm}
\hspace{1.8cm}
\includegraphics[scale=0.5]{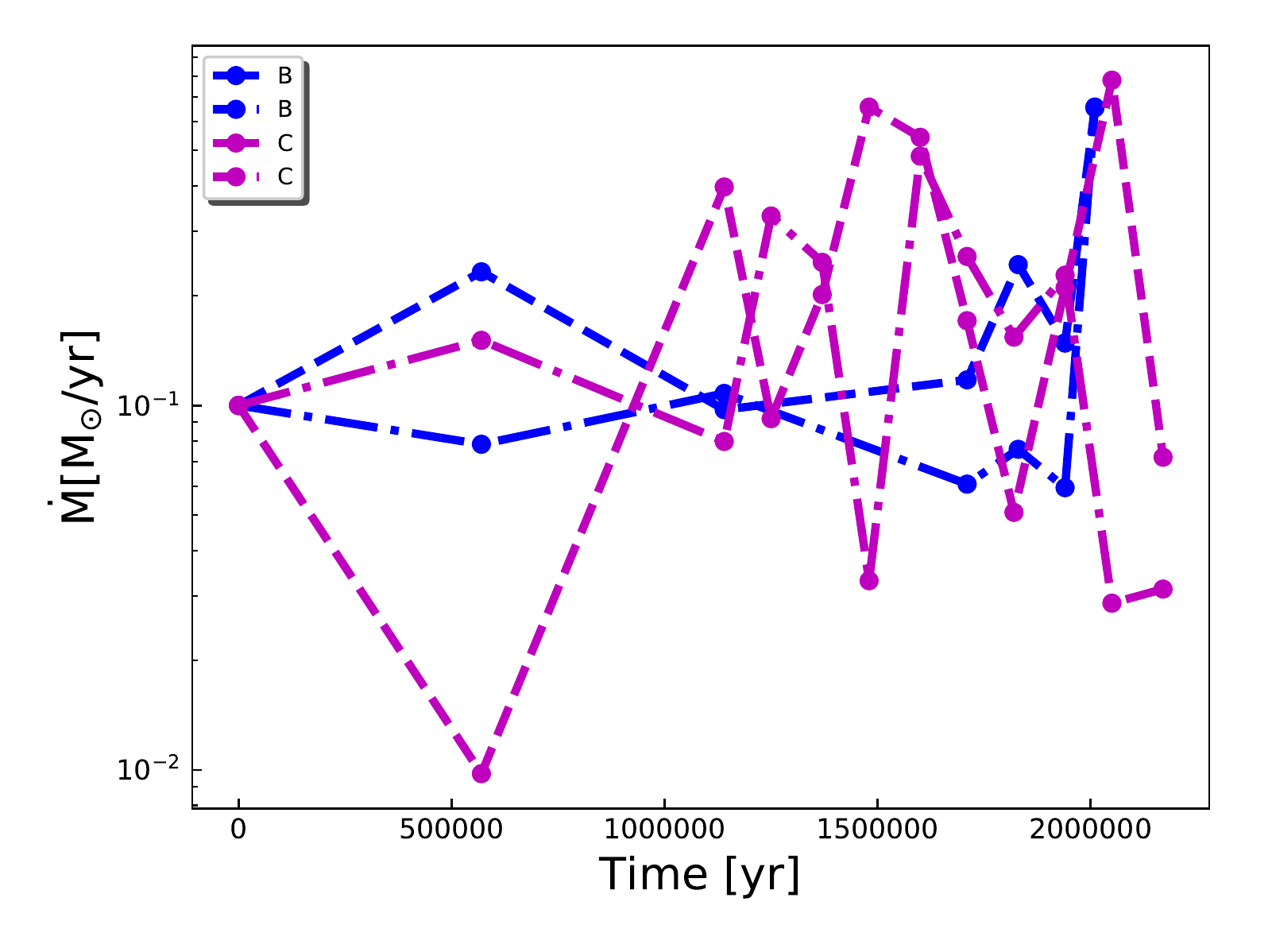}
\end{minipage} \\
\begin{minipage} {6cm}
\vspace{-0.2cm}
\includegraphics[scale=0.5]{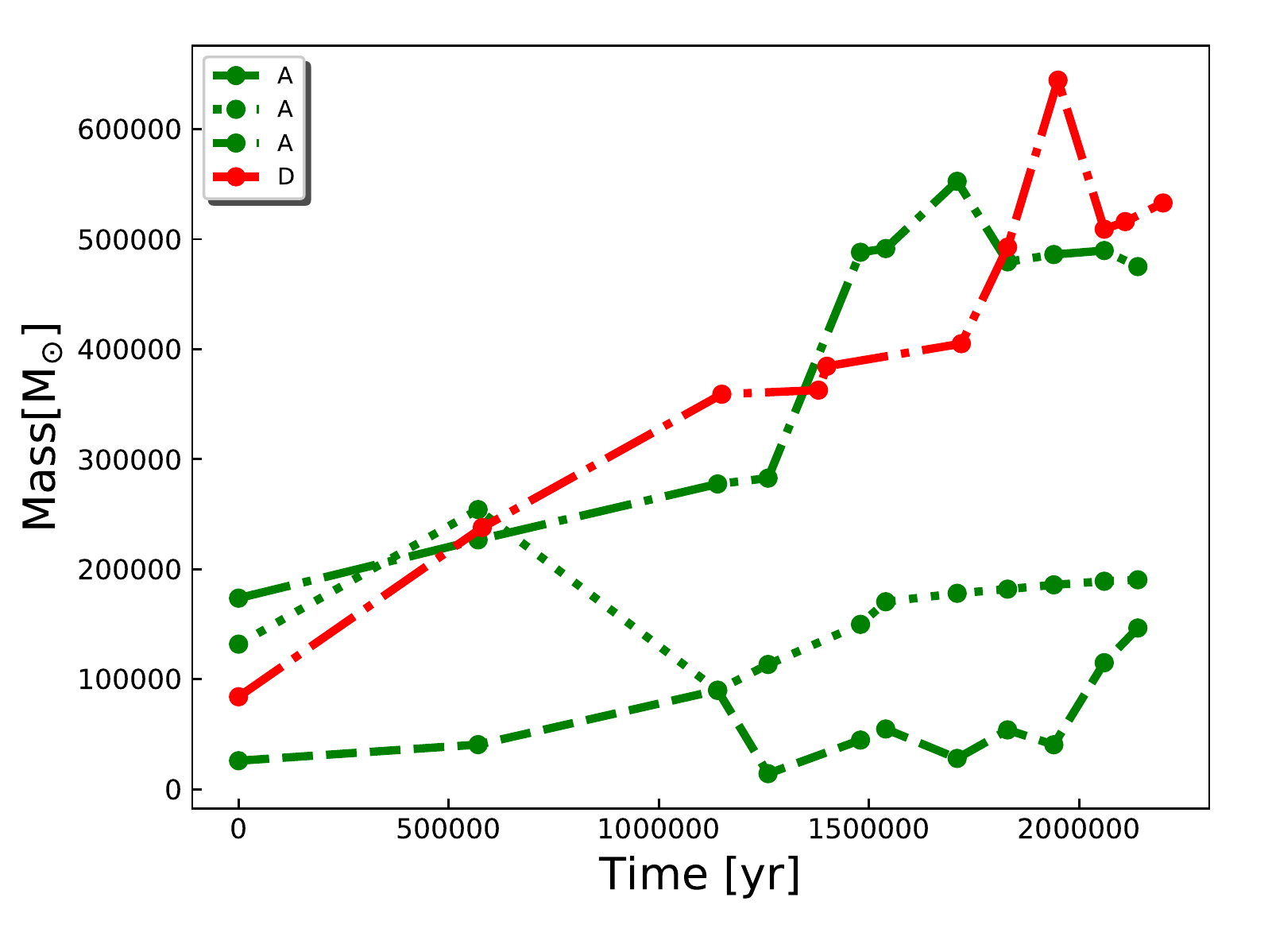}
\end{minipage} &
\begin{minipage}{6cm}
\hspace{1.8cm}
\includegraphics[scale=0.5]{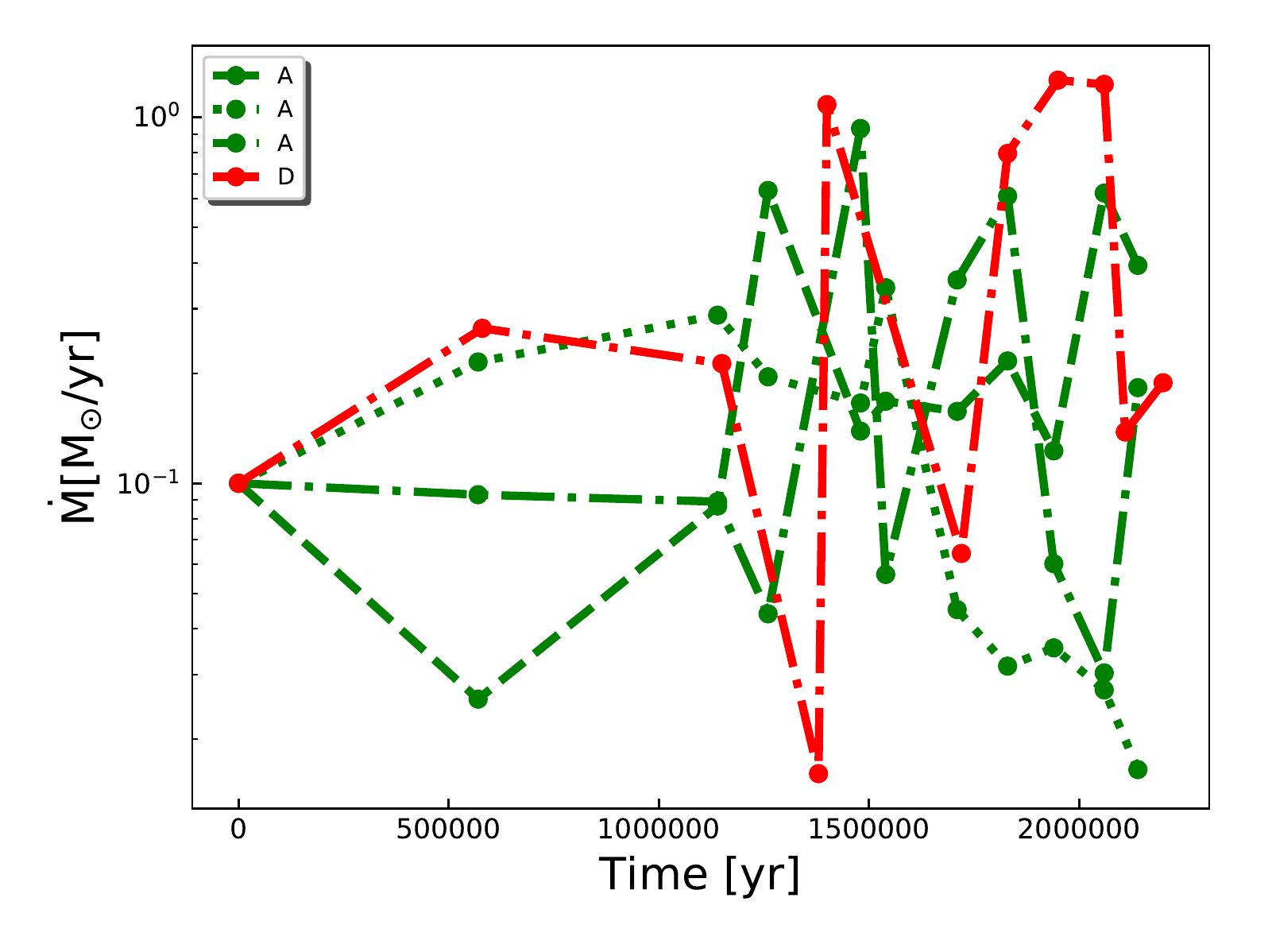}
\end{minipage}
\end{tabular}
\caption{Clump mass vs. time (left panels) and clump accretion rate (right panels).  Upper panels: halos B (blue) and C (magenta).  Lower panels: halos A (green) and D (red). Only clumps surviving for more than 1 Myr are shown.}
\label{fig4}
\end{figure*}

We plot the masses of clumps that survive for more than 1 Myr along with their accretion rates in Figure~\ref{fig4}.  Not all the clumps forming in the high spin halos appear because some already merged with the central object, while others only form in the last 0.6 Myr. The fluctuations in the accretion rates are due to fragmentation and highly turbulent flows in the disks.  It is clear that the large inflow rates previously suspected (but never confirmed) to build up SMSs indeed persist down to small enough scales for long enough times to create such stars, and later DCBHs.  The ratios of the masses of the two clumps in the binary systems is initially $\rm M_1/M_2=2/5$ but the smaller ones grow to nearly the same masses as their partners by $\sim$ 1 Myr. Accretion rates in the binary disks average about 0.1 \Ms\ yr$^{-1}$ for two Myr, more than enough time for SMSs to form and collapse to DCBH binaries that could later merge into a single object.  

Even more stars will form in halo A, in which three massive clumps have grown for about 2 Myr at average rates of $\sim$ 0.1 - 0.2 \Ms\ yr$^{-1}$.  One reaches a final mass of 4-5 $\times 10^{5}$ \Ms\ while the other two  grow to 1 - 2 $\times 10^5$ \Ms.  In halo D all the clumps migrate inward and merge with the one at the center, likely forming just one SMS.  It grows to $5 \times 10^5$ \Ms\ by the end of the run.  Although multiple clumps appear in this halo at later times, an SMS will already have formed at its center and collapsed by then, so it is unclear if they can become stars when exposed to X-rays from a DCBH.  Most of the clumps lose mass at times in their evolution because of tidal stripping by other fragments.  At such times the magnitudes of their accretion rates are plotted in the panel on the right since negative rates cannot appear on logarithmic scales.

\section{Conclusions} \label{sec:conc}

Our simulations confirm for the first time that  SMSs, and therefore DCBHs, can form in binaries or even small multiples in halos with low spins and high spins, respectively.  They also demonstrate that flow rates in these halos continue for sufficient long times on small scales to create such objects.  Although fragmentation has been reported in some recent studies they could not follow the evolution of the clumps to determine if they later formed stars or were simply subsumed into the center of the disk.  Our simulations suggest that binary SMSs preferentially form in low-spin halos because of their lower angular momenta while high-spin halos favor the formation of small multiples because the accretion disk breaks up more easily.  
We note that past numerical simulations exhibit fragmentation on much smaller AU scales \citep{Becerra2014} that are not resolved here.  However, these fragments later merge with the central object on timescales of $\sim$ 10 yr and do not become stars themselves.  Our failure to resolve them therefore does not alter the results of our study.

Although we do not include radiative feedback from SMSs in our simulations, it is not expected to have a large effect on the evolution of the clumps over time.  \citet{Chon18} examined the impact of UV feedback from SMSs in the unlikely scenario that they become blue, hot and luminous in ionizing UV.  They found that the stars created bipolar \ion{H}{2} regions in which the temperature of the gas was at most twice that of the surrounding gas.  Mass loading by infall halts the expansion of the I-front and confines the ionized gas  onto the disk around  the star, with no effect on its growth.  However, \citet{aaron17} post processed highly-resolved simulations of atomically cooling halos with Ly$\alpha$ photon transport and found it might exert some mechanical feedback on flows in the vicinity of the star.  Radiation hydrodynamical simulations by \citet{luo18} and \citet{ard18} without resonant Ly$\alpha$ scattering found that radiation from the protostar in its early stages did not significantly alter flows in its vicinity but did suppress fragmentation, thus promoting the rapid growth of a single supermassive object, but they only evolved these systems for a few years and could not evaluate its effects at later times.

Our simulations have important consequences for the detection of DCBHs (and thus the first quasars) at birth.  Inspirals of two or more DCBHs could emit gravitational wave signals that are powerful enough to be detected at high redshifts by LISA.  Likewise, if the environments of multiple DCBHs are dense with other fragments or less massive (and longer-lived) SMSs, they could produce tidal disruption events (TDEs) that would be extremely luminous in the NIR today \citep{ki16}.  They could be found by the {\em James Webb Space Telescope} ({\em JWST}) or extremely large telescopes (ELTs) in the coming decade.

\section{Acknowledgements}

MAL thanks the UAEU for funding via UPAR grant No. 31S390 and startup grant No 31S372. 
D.~J.~W. was supported by the Ida Pfeiffer Professorship at the Institute of Astrophysics at the University of Vienna and by STFC New Applicant Grant ST/P000509/1.  

\bibliography{smbhs}
\bibliographystyle{aasjournal}

\end{document}